\documentclass[superscriptaddress, 12pt, reprint]{revtex4-1}
\usepackage{amsmath,amssymb,graphicx,color}
\usepackage{soul}

\begin{document}

\title{Digital video microscopy enhanced by deep learning}

\author{Saga Helgadottir}
\affiliation{Department of Physics, University of Gothenburg, 41296 Gothenburg, Sweden, EU}

\author{Aykut Argun}
\affiliation{Department of Physics, University of Gothenburg, 41296 Gothenburg, Sweden, EU}

\author{Giovanni Volpe}
\affiliation{Department of Physics, University of Gothenburg, 41296 Gothenburg, Sweden, EU}

\date{\today}

\begin{abstract}
Single particle tracking is essential in many branches of science and technology, from the measurement of biomolecular forces to the study of colloidal crystals.
Standard current methods rely on algorithmic approaches: by fine-tuning several user-defined parameters, these methods can be highly successful at tracking a well-defined kind of particle under low-noise conditions with constant and homogenous illumination.
Here, we introduce an alternative data-driven approach based on a convolutional neural network, which we name DeepTrack.
We show that DeepTrack outperforms algorithmic approaches, especially in the presence of noise and under poor illumination conditions.
We use DeepTrack to track an optically trapped particle under very noisy and unsteady illumination conditions, where standard algorithmic approaches fail.
We then demonstrate how DeepTrack can also be used to track multiple particles and non-spherical objects such as bacteria, also at very low signal-to-noise ratios.
In order to make DeepTrack readily available for other users, we provide a Python software package, which can be easily personalized and optimized for specific applications.
\end{abstract}

\maketitle

\section*{Introduction}

In many experiments, the motion of a microscopic particle is used as a local probe of its surrounding microenvironment. 
For example, it is used to calibrate optical tweezers \cite{jones2015optical}, to measure biomolecular forces \cite{neuman2008single}, to explore the rheology of complex fluids \cite{waigh2016advances}, to monitor the growth of colloidal crystals \cite{li2016assembly}, and to determine the microscopic mechanical properties of tissues \cite{sahoo2017nanoparticles}.
The first necessary step in order to be able to statistically analyze the trajectory of a microscopic particle is to track the particle position.
This is often done by acquiring a video of the particle and then employing computer algorithms to determine the particle position frame by frame.
This technique was introduced about 20 years ago and is generally referred to as digital video microscopy \cite{crocker1996methods}.
With the increasingly improved quality of digital image acquisition devices and the steady growth in available computational power, the experimental bottleneck has now become the determination of the best data analysis algorithm and its relative parameters for each specific experimental situation. 

Currently, single particle tracking is dominated by algorithmic approaches.
In fact, a large number of algorithms have been developed, especially to track fluorescent particles and molecules \cite{crocker1996methods, sbalzarini2005feature, rogers2007precise, andersson2008localization, manley2010single, parthasarathy2012rapid, thompson2002precise}. 
Some of the most commonly employed algorithms are
the calculation of the centroid of the particle after thresholding the image to convert it to black and white \cite{crocker1996methods},
and the calculation of the radial symmetry center of the particle \cite{parthasarathy2012rapid}. 
When their parameters and the image acquisition process are optimized by the user, these methods routinely achieve subpixel resolution.
However, because of their algorithmic nature, they perform best when their underlying assumptions are satisfied; in general, these assumptions are that the particle is spherically symmetric, that the illumination is homogenous and constant, and that the particle remains in the same focal plane for the whole duration of the experiment.
Their performance degrades severely at low signal-to-noise ratios (SNRs) or under unsteady or inhomogeneous illumination, often requiring significant intervention by the user to reach an acceptable performance, which in turn introduces user bias.
In practice, in these conditions, scientists need to manually search the space of available algorithms and parameters, a process that is often laborious, time-consuming, and user-dependent.
This has severely limited the widespread uptake of these methods outside specialized research labs, while leading to a flourishing of research activity devoted to comparing their performance in challenging conditions \cite{cheezum2001quantitative, ober2004localization, abraham2009quantitative, chenouard2014objective}.

Alternative, data-driven approaches have been comparatively largely overlooked, despite their potential for better, more autonomous performance. 
In fact, data-driven deep learning algorithms based on convolutional neural networks \cite{lecun2015deep} have been extremely successful in image recognition and classification for a wealth of applications from face recognition \cite{parkhi2015deep} to microscopy \cite{litjens2017survey}, and some pioneering work has already shown some of their capabilities for particle tracking \cite{newby2018convolutional, hannel2018machine}. 

Here, we introduce a fully automated deep-learning network design that achieves subpixel resolution for a broad range of particle kinds, also in the presence of noise and under poor, unsteady illumination conditions.
We demonstrate this approach tracking an optically trapped particle under very noisy and unsteady illumination conditions, where standard algorithmic approaches fail.
Then, we show how it can be used to track multiple particles and non-spherical objects such as bacteria.
In order to make this approach readily available for other users, we provide a Python software package, called DeepTrack, which can be readily personalized and optimized for the needs of specific users and applications.

\section*{Results}

\subsection*{DeepTrack neural network architecture and performance}

\renewcommand{\figurename}{{\bf \footnotesize Figure}}
\setcounter{figure}{0}
\renewcommand{\thefigure}{{\bf \arabic{figure}}}
\begin{figure*}[ht!]
\includegraphics [width=0.7\textwidth]{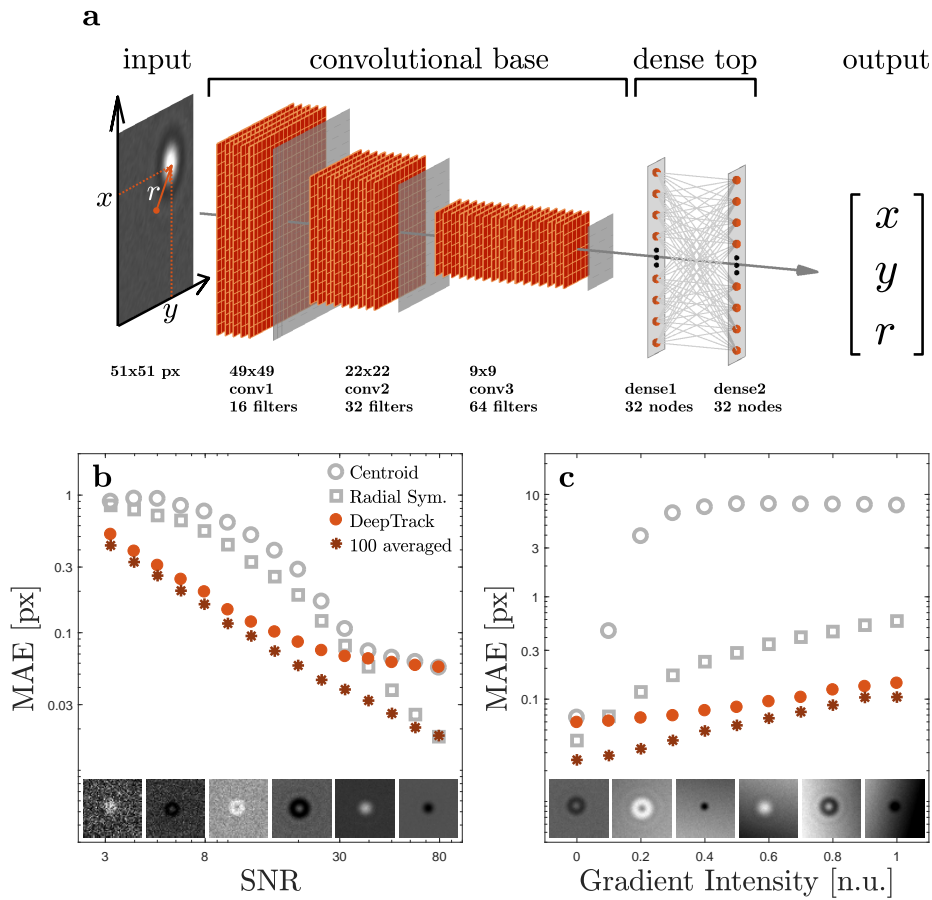}
\caption{\footnotesize
{\bf DeepTrack neural network architecture and performance.}
{\bf a} DeepTrack architecture consists of a convolutional base (3 convolutional neural network layers, depicted in orange, each followed by a max-pooling layer, depicted in gray) followed by a dense top (two fully connected dense layers and a dense output layer). 
In the convolutional base, the image is iteratively filtered to extract an increasing number of feature maps and downsampled. 
In the dense top, the feature maps are used to predict the values of the x-, y-, and r-coordinates of the particle.
{\bf b} Mean absolute error (MAE) of the position detection as a function of signal-to-noise ratio (SNR) for DeepTrack (orange circles) and the centroid (gray circles) and radial symmetry (gray squares) algorithms. 
The purple stars represent the performance achieved by averaging the coordinates obtained with 100 independently-trained neural networks.
{\bf c} Same as {\bf b} as a function of the gradient intensity at SNR=50.
For each SNR or gradient intensity value we used 1000 simulated images; the error bars are contained within the symbols.
See also Example Code 1a, which demonstrates the performance of a pre-trained neural network, and Example Code 1b, which illustrates the training and operation of the neural network.
}
\label{Fig1}
\end{figure*}

While standard algorithmic approaches require the user to give explicitly rules (i.e. the algorithm) to process the input data in order to obtain the sought-after result, 
machine learning systems are trained through large series of input data and corresponding results from which they autonomously determine the rules for performing their assigned task.
Neural networks are one of the most successful tools for machine learning \cite{chollet2017deep}: they consist of a series of layers that, when appropriately trained, output increasingly meaningful representations of the input data leading to the sought-after result; these layers can be of various kinds (for example, convolutional layers, max-pooling layers, and dense layers) and their number is the network depth (hence the term deep learning).
In particular, convolutional neural networks have been shown to perform well in image classification \cite{krizhevsky2012imagenet, karpathy2014large, he2015delving} and regression tasks \cite{silver2016mastering}: their architecture consists of a series of convolutional layers (convolutional base) followed by a series of dense layers (dense top).
In each convolutional layer, a series of 2D filters is convolved with the input image, producing as output a series of feature maps.
The size of the filters with respect to the input image determines the features that can by detected in each layer: to gradually detect larger features, the feature maps are downsampled by adding a max-pooling layer after each convolutional layer. The max-pooling layers retain the maximum values of the feature maps over a certain area of the input image. The downsampled feature maps are then fed as input to the next network layer.  
After the last convolutional layer there is a dense top, which consists of fully connected dense layers. These layers integrate the information contained in the output feature maps of the last max-pooling layer to determine the sought-after result.
Initially the weights of the convolutional filters and of the dense layers are random, but they are iteratively optimized using the back-propagation algorithm \cite{rumelhart1986parallel}. 

A schematic of the neural network architecture we employed in this work is shown in Fig.~\ref{Fig1}{\bf a} and its details are provided in the Methods.
Given an input image, this neural network returns the x-, y-, and r-coordinates of the particle, where the x- and y-coordinates are the Cartesian coordinates and the r-coordinate is the radial distance of the particle from the center of the image. Even though the r-coordinate might seem redundant, it is useful to identify images with no particles, for which the neural network is automatically trained to return a very large r-coordinate value as these images resemble images with a particle that is far outside the frame.
We have implemented this neural network using the Python-based Keras library \cite{chollet2015keras} with a TensorFlow backend \cite{tensorflow2015-whitepaper} because of their broad adoption in research and industry; nevertheless, we remark that the approach we propose is independent of the deep learning framework used for its implementation.

Once the network architecture is defined, we need to train it on a set of particle images for which we know the ground-truth values of the x-, y-, and r-coordinates of the particle.
In each training step, the neural network is tasked with predicting the coordinates corresponding to a set of images; the neural network predictions are then compared to the ground-truth values of the coordinates; and the prediction errors are finally used to adjust the trainable parameters of the neural network using the back-propagation training algorithm \cite{rumelhart1986parallel}.
The training of a neural network is notoriously data intensive, requiring in our case several millions particle images, as described in detail in the Methods. Therefore, in order to have enough images and to accurately know the ground-truth values of the corresponding coordinates, we simulate the particle images. 
The particle generation routine is described in detail in the Methods; briefly, we generate the images of the particles using Bessel functions because they very well approximate the appearance of colloidal particles in digital video microscopy \cite{bohren2008absorption}.
Setting the parameters of the image generation function, we can generate images where the particles are represented by dark or bright spots or rings of different intensities on a bright or dark background with varying SNR and illumination gradients; some examples of these images can be seen in the insets of Figs.~\ref{Fig1}{\bf b}-{\bf c}.
We train the neural network using a grand total of about $1.5$-million images, which we present to the network in gradually increasing batches (see Methods); in this way, at the beginning the neural network optimization process can easily explore a large parameter space and later it gets annealed towards an optimal parameter set \cite{smith2017don}.
We simulate a new batch of images before each training step. Since each image is shown to the network only once and then discarded, this permits us to make a very efficient use of the computer memory as well as, more importantly, to prevent overtraining and to avoid the need for real-time validation of the network performance.
Overall, this training process is very efficient, taking about 3 hours on a standard laptop computer, which can be reduced by up to two orders of magnitude on a GPU-enhanced computer.
Furthermore, we remark that once the neural network is trained, its use is very computationally efficient and its execution time is comparable to that of standard algorithms.

In Fig.~\ref{Fig1}{\bf b}, we test the performance of DeepTrack on simulated images with a range of SNR values and homogeneous illumination (i.e. without any illumination gradient). Examples of these images are shown in the insets of Fig.~\ref{Fig1}{\bf b} with increasing SNR from left to right.
As shown by the orange circles in Fig.~\ref{Fig1}{\bf b}, DeepTrack achieves subpixel accuracy over all the range of SNRs, from SNR=3.2 (noisiest images on the left) to SNR=80 (almost perfect images on the right).
We then benchmark DeepTrack against the centroid (gray circles) and radial symmetry (gray squares) algorithms, by testing them on the same set of images. While the radial symmetry algorithm is better for almost perfect images, DeepTrack outperforms both algorithms in high-noise conditions up to SNR=40 and the centroid algorithm over the whole range of SNRs.
Furthermore, the performance of DeepTrack can be significantly improved averaging the particle coordinates obtained from several independently-trained neural networks; for example, the purple stars in Fig.~\ref{Fig1}{\bf b} represent the results obtained averaging 100 neural networks and show that the neural-network approach outperforms both algorithmic approaches over the whole range of SNRs.

In Fig.~\ref{Fig1}{\bf c}, we explore the influence of illumination gradients, which are known to introduce artefacts in digital video microscopy.
In fact, illumination gradients are often present in experiments as a consequence of illumination inhomogeneities, for example, due to the presence of out-of-focus microfluidic structures or biological tissues. 
Although it is sometimes possible to correct for such gradients by image flattening, often the direction and intensity of the gradient varies as a function of position and time so that it cannot be straightforwardly corrected \cite{rogers2007precise}.
We test DeepTrack on $1000$ SNR=50 images affected by a range of illumination gradients with random direction. Examples of these images are shown in the insets of Fig.~\ref{Fig1}{\bf c} with increasing intensity gradient from left to right.
As shown by the orange circles in Fig.~\ref{Fig1}{\bf c}, DeepTrack predicts the particle coordinates with an accuracy of less than $0.1$ pixels over the all range of gradient intensities, in contrast to the performances of the centroid (gray circles) and radial symmetry (gray squares) algorithms that rapidly deteriorate as soon as any intensity gradient is present.
As shown by the purple stars, also in this case, the performance of DeepTrack can be significantly improved by averaging the coordinates obtained from multiple independently-trained neural networks.

\subsection*{Experimental tracking of an optically trapped particle}

\renewcommand{\figurename}{{\bf \footnotesize Figure}}
\setcounter{figure}{1}
\renewcommand{\thefigure}{{\bf \arabic{figure}}}
\begin{figure*}[ht!]
\includegraphics [width=1.0\textwidth]{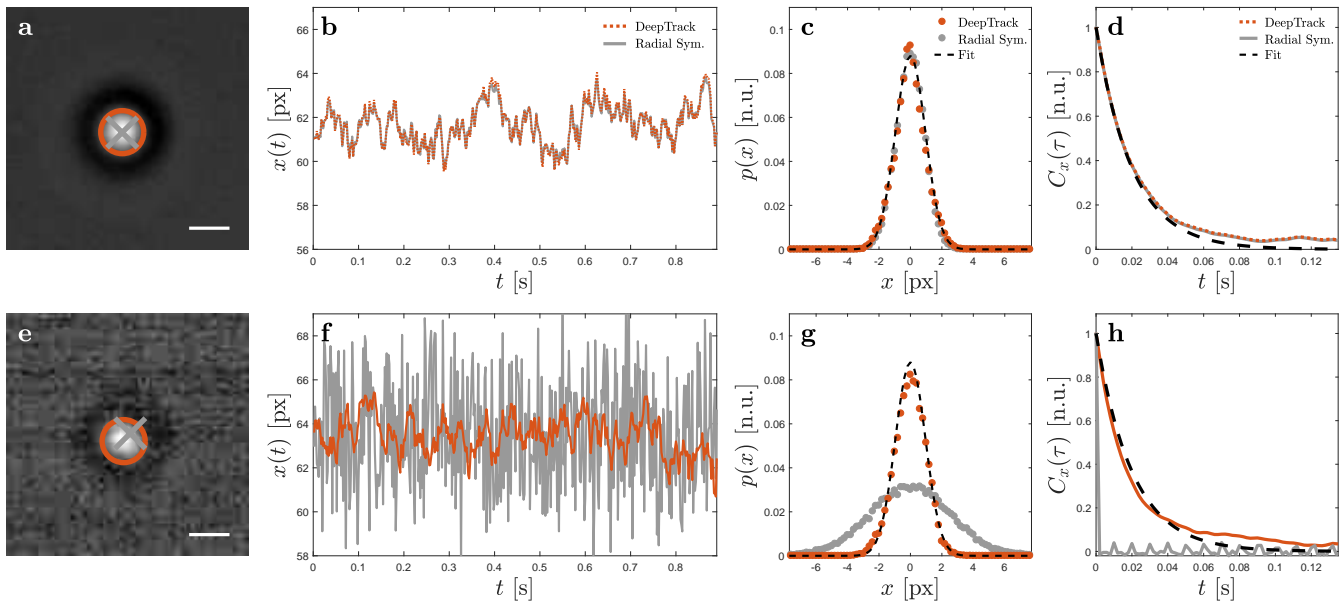}
\caption{\footnotesize
{\bf Experimental tracking of an optically trapped particle.}
{\bf a}-{\bf d} DeepTrack and standard algorithms lead to the same results when tracking and analysing the trajectory of an optically trapped particle (silica microsphere, diameter $1.98\,{\rm \mu m}$) under optimal illumination conditions (see also Video 1): 
{\bf a} Image of the optically trapped particle, and its position obtained by DeepTrack (orange circle) and by the radial symmetry algorithm (gray cross); 
{\bf b} Part of the trajectory tracked by DeepTrack (orange line) and by the radial symmetry algorithm (gray line); 
{\bf c} Probability distributions and {\bf d} autocorrelation functions of the particle position obtained from the trajectory tracked by DeepTrack (orange lines) and by the radial symmetry algorithm (gray lines), and corresponding fittings to theory (black lines).
{\bf e}-{\bf h} DeepTrack outperforms standard algorithms when the illumination is unsteady (here obtained by illuminating the sample with a standard lamp flickering at $100\,{\rm Hz}$) and noisy (setting the camera gain to its highest value):
{\bf e} Image of the same particle in the same optical trap with noisy illumination (see also Video 2);
{\bf f} The trajectory reconstructed by DeepTrack appear qualitatively more similar to those shown in {\bf b};
{\bf g} The probability distribution and {\bf h} the autocorrelation function obtained by DeepTrack (orange lines) agree with those obtained in {\bf c} (the black lines are the same as in {\bf c} and {\bf d}), while the probability distribution from the radial symmetry algorithm (gray symbols in {\bf g}) is widened by the noise, and the autocorrelation function (gray line in {\bf h}) features a large peak at $0$, which is the signature of a white noise, and some small oscillations at $100\,{\rm Hz}$, which are due to the flickering of the illumination.
The scale bars in {\bf a} and {\bf e} represent 20 pixels corresponding to $1.4\,{\rm \mu m}$.
See also Example Code 2, which uses a pre-trained neural network to track these particles.
}
\label{Fig2}
\end{figure*}

Until now we have trained and tested our neural network on simulated images. In order to see how DeepTrack works in real life, we now test its performance on some experimental images, while still training it on simulated images as above.
As a first experiment, we track the trajectory of a particle held in an optical tweezers (see Methods). An optical tweezers is a focused laser beam that can trap a microscopic particle near its high-intensity focal spot \cite{jones2015optical}. From the motion of the particle, it is possible to calibrate the optical tweezers and then use it for quantitative force measurement. The critical step is to track the position of the particle, which is often done using digital video microscopy.
We record the video of an optically trapped microsphere (silica, diameter $1.98\,{\rm \mu m}$) in an optical tweezers (wavelength $532\,{\rm nm}$, power $2.5\,{\rm mW}$ at the sample).
First, we record the particle video illuminating the sample with a high-power LED, which offers ideal illumination conditions and, therefore, provides us with a very clear image of the particle (Fig.~\ref{Fig2}{\bf a} and Video 1). In these conditions, standard methods (we show results for the radial symmetry algorithm because it is the one that performs best amongst the standard methods we tested) can track the particle extremely well (gray cross in Fig.~\ref{Fig2}{\bf a} and Video 1, and gray line in Fig.~\ref{Fig2}{\bf b}) and serve as a benchmark for our neural network (orange circle in Fig.~\ref{Fig2}{\bf a} and Video 1, and orange line in Fig.~\ref{Fig2}{\bf b}): the two trajectories agree to within 0.089 pixels (mean absolute difference).
In order to obtain a more quantitative measure of the agreement between the two tracking methods, we calculated the probability distribution (Fig.~\ref{Fig2}{\bf c}) and the autocorrelation function (Fig.~\ref{Fig2}{\bf d}) of the particle position, which are standard methods to calibrate optical tweezers \cite{jones2015optical}: in both cases, the DeepTrack results (orange lines) agree well with the standard algorithm results (gray lines); furthermore, these results agree with the fits to the corresponding theoretical functions (black lines), that are respectively a Gaussian function (see Methods) and an inverted exponential (see Methods).
Here, we trained DeepTrack with about 1.5-million simulated images similar to the experimental images, where the particle is represented by the sum of a Bessel function of first order with positive intensity (bright spot) and a Bessel function of second order with negative intensity (dark ring), and the background level, SNR and illumination gradient are randomized for each image (see Example Code 2).

We now make the experiment more challenging by substituting the LED illumination with a very poor illumination device: a low-power incandescence light bulb connected to an AC plug placed $10\,{\rm cm}$ above the sample without any condenser. The $50$-${\rm Hz}$ AC current results in the illumination light flickering at $100\,{\rm Hz}$, and the low power requires us to increase to the maximum the gain of the camera leading to a high level of electronic noise (Fig.~\ref{Fig2}{\bf e} and Video 2).
Even in these very challenging conditions, DeepTrack manages to track the particle position accurately (orange circle in Fig.~\ref{Fig2}{\bf e} and Video 2, and orange line in Fig.~\ref{Fig2}{\bf f}), while the standard tracking algorithm loses its accuracy (gray cross in Fig.~\ref{Fig2}{\bf e} and Video 2, and gray line in Fig.~\ref{Fig2}{\bf f}).
We can quantify these observations by calculating the probability distribution (Fig.~\ref{Fig2}{\bf g}) and the autocorrelation function (Fig.~\ref{Fig2}{\bf h}) of the particle position.
The probability distribution calculated from the trajectory obtained by the standard method (gray line in Fig.~\ref{Fig2}{\bf g}) is significantly widened because of the presence of the illumination noise, while that from the DeepTrack trajectory (orange line) agrees well with the theoretical fit (black line, same as in Fig.~\ref{Fig2}{\bf c}).
Even more strikingly, the autocorrelation from the standard algorithm trajectory (gray line in Fig.~\ref{Fig2}{\bf h}) does not retain any of the properties of the motion of an optically trapped particle: it features a large peak at $\tau=0$, which is the signature of a white noise, and some small oscillations at 100 Hz, which are due to the flickering of the illumination. Instead, the autocorrelation from the DeepTrack trajectory (orange line) agrees very well with the theoretical fit (black line, same as in Fig.~\ref{Fig2}{\bf d}), demonstrating that the neural network successfully removes essentially all the overwhelming noise introduced by the poor illumination, removing also the heavy flickering of the light source as shown by the absence of 100-Hz oscillations.

\subsection*{Tracking of multiple particles}

\renewcommand{\figurename}{{\bf \footnotesize Figure}}
\setcounter{figure}{2}
\renewcommand{\thefigure}{{\bf \arabic{figure}}}
\begin{figure*}[ht!]
\includegraphics [width=1.0\textwidth]{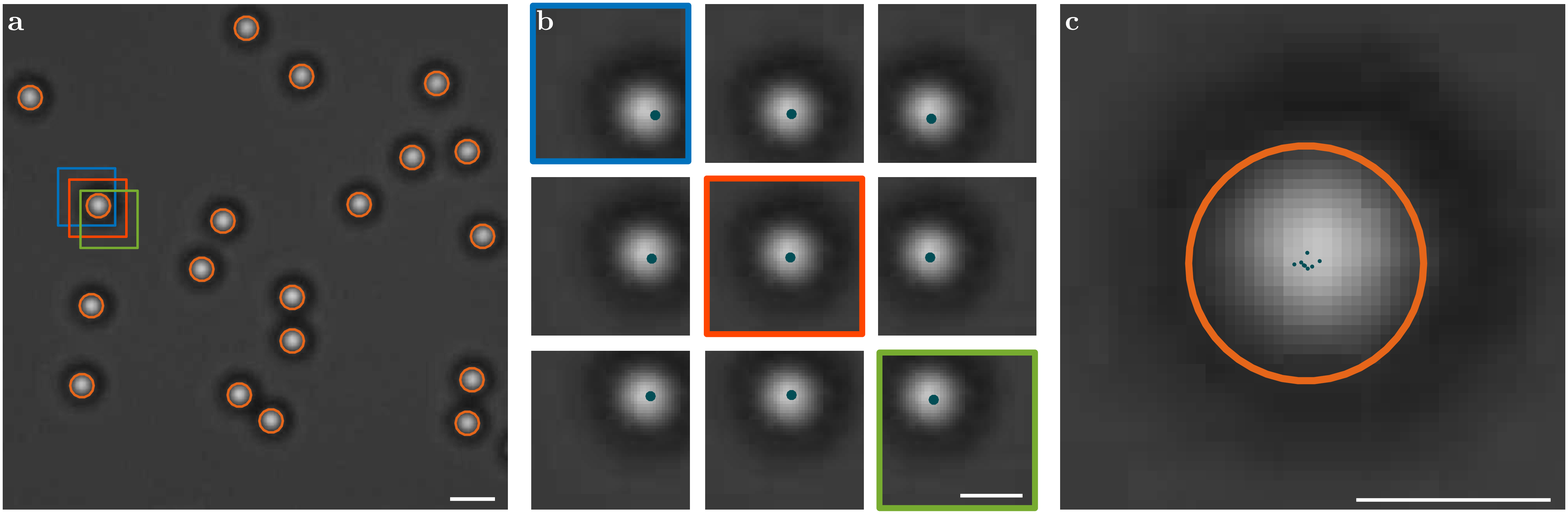}
\caption{\footnotesize
{\bf Tracking of multiple particles.}
{\bf a} DeepTrack tracks the position (orange circles) of several microspheres (silica, diameter $1.98\,{\rm \mu m}$) diffusing above a coverslip (see also Video 3 and Example Code 3). 
The tracking is performed as follows:
First, the frame is divided into overlapping square boxes (for example, the  blue, red and green boxes) whose side is approximatively twice the particle diameter (here we use $51\times51$-pixel boxes separated by $5$ pixels).
{\bf b} Then, each box is tracked and the particle x-, y-, and r-coordinates are determined (the blue, red and green boxes correspond to those shown in {\bf a}), so that each particle is detected multiple times (blue dots in each square); if no particle is present within a box, the network is trained to return a value of the radial distance much larger than the particle radius; importantly, only the particle coordinates for which the radial coordinate is smaller than the particle diameter are retained.
{\bf c} Finally, the multiple particle detections are clustered into sets of points whose inter-distance is smaller than the particle radius (blue dots) and the corresponding particle x- and y-coordinates are then obtained by calculating the centroid of these points (orange circle). 
All scale bars indicate 20 pixels corresponding to $1.4\,{\rm \mu m}$.
}
\label{Fig3}
\end{figure*}

\renewcommand{\figurename}{{\bf \footnotesize Figure}}
\setcounter{figure}{3}
\renewcommand{\thefigure}{{\bf \arabic{figure}}}
\begin{figure*}[ht!]
\includegraphics [width=1.0\textwidth]{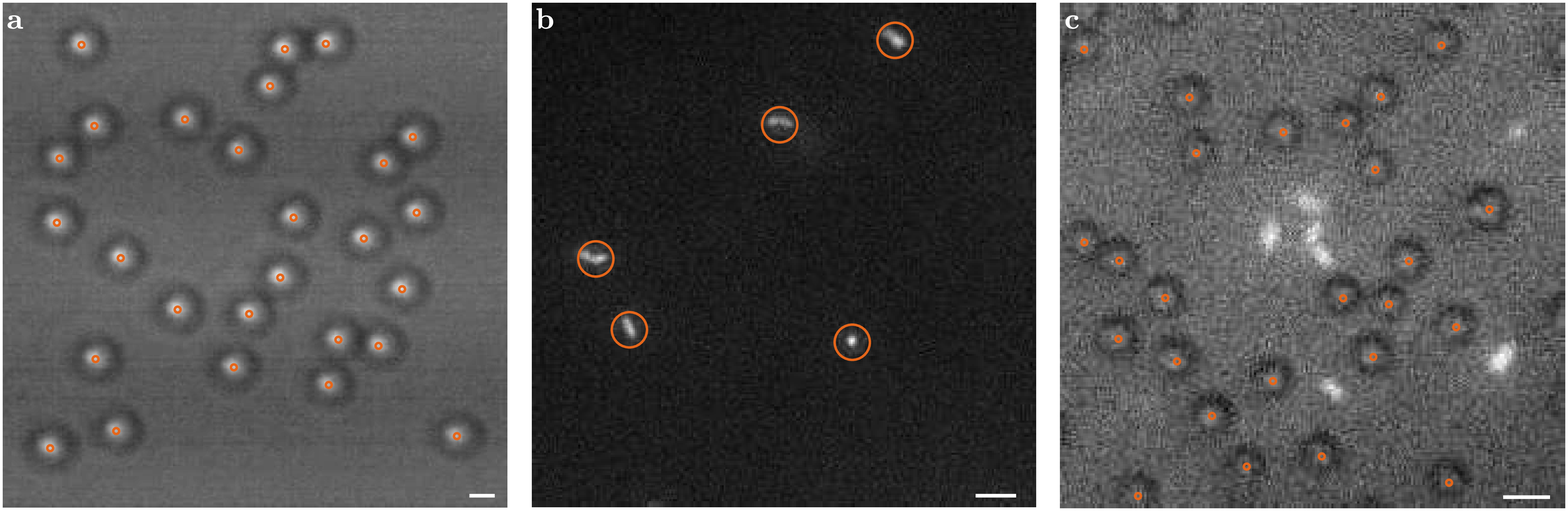}
\caption{\footnotesize
{\bf Tracking of multiple particles and bacteria in poor illumination conditions.}
Using the same training as in Fig.~\ref{Fig3}, DeepTrack manages to determine the positions (orange circles) 
{\bf a} of the same Brownian particles as in Fig.~\ref{Fig3} but in unsteady, non-uniform illumination conditions (see also Video 4 and Example Code 3),
and {\bf b} of \emph{B. subtilis} bacteria, which are non-spherical objects for which the neural network has not been specifically trained (see also Video 5 and Example Code 3).
{\bf c} With additional training, DeepTrack manages also to track Brownian particles at very low SNR while ignoring fluorescent bacteria present in the background (see also Video 6 and Example Code 4).
All scale bars indicate 20 pixels corresponding to {\bf a} $1.4\,{\rm \mu m}$ and {\bf b}-{\bf c} $5.2\,{\rm \mu m}$.
}
\label{Fig4}
\end{figure*}

In many experiments in, e.g., biology \cite{crocker2007multiple}, statistical physics \cite{berut2014energy}, and colloidal science \cite{baumgartl2005limits}, it is necessary to track multiple particles simultaneously. 
To demonstrate how DeepTrack works in a multi-particle case, we record a video with multiple microspheres (silica, diameter $1.98\,{\rm \mu m}$) under the same ideal illumination condition as in Figs.~\ref{Fig2}{\bf a}-{\bf d}. A sample frame is shown in Fig.~\ref{Fig3}{\bf a} (see also Video 3). 
DeepTrack is used as above with only three modifications (see also Example Code 3).
First, during the training, the neural network is exposed to simulated images containing multiple particles as well as images of single particles, and it is in all cases trained to determine the coordinates of the most central particle in each image (here, we employed 2 million images containing one to four particles similar to those employed in Fig.~\ref{Fig2}).
Second, during the detection, the frame is divided into overlapping square boxes (for example, the  blue, red and green boxes shown in Fig.~\ref{Fig3}{\bf a}) whose side is approximatively twice the particle diameter (in Fig.~\ref{Fig3}, we use $51\times51$-px boxes separated by $5$ pixels), and for each box the neural network determines the coordinates of the most central particle (e.g., blue dots in Fig.~\ref{Fig3}{\bf b}); importantly, in order to easily detect empty boxes, the neural network is trained to return a large radial distance if no particle is present; therefore, only particle positions for which the radial distance is smaller than a certain threshold (typically, a value between particle radius and diameter) are retained for further analysis.
Third, depending on the size of the boxes and their separation, the same particle can be detected multiple times; therefore, all detections (blue dots in Fig.~\ref{Fig3}{\bf c}) whose inter-distance is smaller than a certain threshold (typically, a value between particle radius and diameter) are assigned to the same particle; and the particle coordinates are then calculated as the centroid of these positions (orange circle in Fig.~\ref{Fig3}{\bf c}). Following this procedure, DeepTrack can accurately track particles even when they are close to contact, which is relevant in many experiments.
There is a trade-off between detection accuracy and computational efficiency, as a smaller separation between the boxes increases the number of detections available to estimate the particle coordinates, but requires more computational resources.
Also, it can be noticed in Video 3 that the particles closer than slightly less than a half-box distance to the border of the image are not detected; if necessary, this effect can easily be eliminated by padding the image.

We can now test DeepTrack on more challenging multi-particle videos, as shown in Fig.~\ref{Fig4} (see also Example Code 3).
Using the same training of the neural network used in Fig.~\ref{Fig3}, we can also detect the same kind of Brownian particles but in the much more challenging illumination conditions already employed in Figs.~\ref{Fig2}{\bf e}-{\bf h}, where the illumination is provided by a low-power, flickering incandescence light bulb (Fig.~\ref{Fig4}{\bf a} and Video 4),
as well as rod-shaped fluorescent bacteria (\emph{Bacillus subtilis}, wild type strain NCIB 3610) which are correctly tracked despite the neural network having been trained only on spherical particles (Fig.~\ref{Fig4}{\bf b} and Video 5).

In addition to the versatile application of a generically trained neural network, we also show how the training of the neural network can be optimized for specific needs. In experiments it is often needed to track only a certain kind of particles while ignoring others. For example, we consider the more challenging case where we want to track only one of two kinds of objects in very noisy conditions. We record a video with microspheres (silica, diameter $4.28\,{\rm \mu m}$) and fluorescent \emph{B. subtilis} bacteria (see Methods) with electronic noise due to the camera gain at low illumination and changes in intensity of the background, particles and bacteria from frame to frame due to changes in the frame contrast. We extensively tried to track this video in a reliable way using other standard algorithms without success. We train the neural network with simulated images of multiple particles similar to the two kinds present in the solution (a second-order Bessel function with negative intensity for the Brownian particles, and a first-order Bessel function with positive intensity for the bacteria) with a low background level and SNR; the ground-truth values given to the neural network in this case are the coordinates for the most central particle in the image, which is set to always be a particle representing a microsphere (thus training to ignore the surrounding bright spots); training with 4 million images allows us to accurately and selectively track only the Brownian particles, as shown in Fig.~\ref{Fig4}{\bf c} (see also Video 6 and Example Code 4).

\section*{Discussion}

We provide a data-driven neural-network approach that  enhances digital video microscopy going beyond the state of the art available through standard algorithmic approaches to particle tracking.
We provide a Python freeware software package, called DeepTrack, which can be readily personalized and optimized for the needs of specific users and applications. 
We have shown that DeepTrack can be trained for a variety of experimentally relevant images and outperforms traditional algorithms especially when image conditions become non-ideal due to low or non-uniform illumination. 
To facilitate adoption of this approach, we provide example codes of the training and testing of DeepTrack for each of the cases we discuss. 
Even though DeepTrack is already sufficiently computationally efficient to be trained on a standard laptop computer in a few hours, its speed can be significantly enhanced taking advantage of parallel computing and GPU-computing, for which neural network are particularly suited \cite{schmidhuber2015deep,chetlur2014cudnn}, especially since the underlying neural-network engine provided by Keras and TensorFlow is already GPU-compatible \cite{chollet2015keras, tensorflow2015-whitepaper}.
Furthermore, this approach can potentially be implemented in the future in hardware \cite{farabet2011large} or using alternative neural-network computing paradigms such as reservoir computing \cite{abadi2016tensorflow}.

\section*{Methods}

\textbf{Neural network architecture.} 
The neural network architecture is schematically shown in Fig.~\ref{Fig1}{\bf a}. 
It consists of a series of convolutional neural network layers followed by max-pooling layers (convolutional base) followed by a series of dense neural network layers (dense top). 
In the convolutional base, the image is iteratively filtered by the convolutional layers using $3\times3$ filters to extract an increasing number of feature maps and downsampled by the max-pooling layers over areas corresponding to $2\times2\,{\rm px}$: 
given that the original image is $51\times51\,{\rm px}$, the first convolutional layer produces 16 feature maps of $49\times49\,{\rm px}$ and the first max-pooling layer downsamples them to $24\times24\,{\rm px}$; 
the second convolutional layer produces 32 feature maps of $22\times22\,{\rm px}$ and the second max-pooling layer downsamples them to $11\times11\,{\rm px}$; 
and the third convolutional layer produces 64 feature maps of $9\times9\,{\rm px}$ and the third max-pooling layer downsamples them to $4\times4\,{\rm px}$.
In the subsequent dense top, these 64 $4\times4\,{\rm px}$ feature maps are fed into two dense layers consisting of 32 nodes each with rectified linear activation, and an output layer of $3$ nodes with linear activation, which perform the final regression task to determine the x-, y-, and r-coordinates of the particle. 
Overall, this results in a neural network with 57251 trainable parameters. We have implemented this neural network using the Python-based Keras library \cite{chollet2015keras} with a TensorFlow backend \cite{tensorflow2015-whitepaper}. 
All the source codes are provided (see below).

\textbf{Neural network training.} 
We train the neural network using simulated images (see below) and the respective ground-truth values for the corresponding x-, y-, and r-coordinates. These images are presented to the neural network in batches.
We gradually increase the batch size during training \cite{smith2017don}, presenting to the network, for example, about 4000, 3000, 2000, 1000 and 500 batches of 8, 32, 128, 512 and 1024 images respectively (grand total of about 1.5-million images).
This training process takes about 3 hours on a standard laptop computer (processor Intel Core i7 at 2.2 GHz and 8 GB 1600 MHz DDR3).

\textbf{Image generation.}
The images of the particles are generated using a point-spread function (PSF) based on a combination of Bessel functions of the first kind of various orders with positive or negative intensity. 
The image generation function accepts the following set of parameters: particle coordinates, particle radius, Bessel function order, particle intensity (which can be either positive or negative), image background, SNR, and illumination gradient.
This permits us to simulate particle images corresponding to dark or bright spots or rings of different intensities on a bright or dark background with varying SNR and illumination gradients.
Some examples of these images can be seen in the insets of Figs.~\ref{Fig1}{\bf b}-{\bf c} and in the Example Codes. 

When generating images of more than one particle, the parameters particle position, radius, Bessel function and intensity are defined for each of the particles. Constrains are added to the particles positions such that they do not overlap and the particle of interest is always closest to the center (see Example Codes 3 and 4).

\textbf{Neural network tracking.}
The images to be tracked are fed as an input to the neural network and for each image the neural network predicts the corresponding x-, y-, and r-coordinates. 
Note that these images do not need to be exactly $51 \times 51$ pixels (the dimension of the input of the first convolutional layer input), as DeepTrack automatically rescales the input images to this size.

When tracking multiple particles, the image is divided into a series of boxes and DeepTrack is used to track the position of the particle in each of these boxes, as described in the main text and Fig.~\ref{Fig3}.
Only particle positions for which the radial distance is smaller than 7.5 (Fig.~\ref{Fig3}), 7.5 (Fig.~\ref{Fig3}{\bf a}), 8.75 (Fig.~\ref{Fig3}{\bf b}) and 3.75 (Fig.~\ref{Fig3}{\bf c}) pixels are retained for further analysis.
Amongst these particle positions, all detections whose inter-distance is smaller than 15 (Fig.~\ref{Fig3}), 25 (Fig.~\ref{Fig3}{\bf a}), 10 (Figs.~\ref{Fig3}{\bf b}-{\bf c}) pixels are assigned to the same particle and, as explained in the main text, the particle position is then calculated as their centroid.
Importantly, note that the final results is not sensitive to the choice of these parameters.

\textbf{Centroid tracking algorithm.} 
The centroid tracking algorithm \cite{crocker1996methods} measures the position of a particle by calculating the centroid of its image after thresholding the image to convert it to black and white. 
We apply this algorithm to each image by subtracting the most frequently occurring gray-scale value, which typically corresponds to the background; taking the absolute value of the resulting image, in order to be able to deal both with bright and dark particle images; and, finally, converting this image into a binary image with a threshold of half of the maximum intensity.

\textbf{Radial symmetry tracking algorithm.} 
The radial symmetry tracking algorithm \cite{parthasarathy2012rapid} localizes an imaged particle center by determining the point of maximal radial symmetry in the image intensity. 
To implement this algorithm, we use the Matlab function radialcenter.m provided in Ref.~\cite{parthasarathy2012rapid}. 

\textbf{Experimental setups.} 
For the results that are shown in Fig.~\ref{Fig2}, we use a standard optical tweezers system built on a home-made microscope \cite{jones2015optical}. We use a dilute solution of microspheres (silica, diameter $1.98\,{\rm \mu m}$, Microparticles GmbH) trapped with a laser beam (wavelength $532\,{\rm nm}$, power $2.5\,{\rm mW}$ at the sample)  focused through a microscope objective (Nikon Plan Fluor 40x Air, NA=0.75). The motion of the optically trapped particle is recorded by a CMOS camera (Thorlabs DCC1645C) at $504\,{\rm fps}$ with a field of view of $120 \times 120\,{\rm px}$ crop ($8.4\times8.4\,{\rm \mu m}$). For the ideal illumination conditions (Fig.~\ref{Fig2}{\bf a}-{\bf d}), we illuminate the sample with a high-power LED (Thorlabs LED MCWHL5, white light, power $38\,{\rm mW}$ at the sample). For the poor illumination conditions (Fig.~\ref{Fig2}{\bf e}-{\bf h}), we substitute the LED illumination with a low-power incandescence light bulb connected to an AC plug ($35\,{\rm W}$, $230\,{\rm lm}$, placed $10\,{\rm cm}$ above the sample without any condenser): the 50-Hz AC current results in the illumination light flickering at 100 Hz, and the low power requires us to increase to the maximum the gain of the camera leading to a high level of electronic noise.

For the multiple-particle experiments shown in Figs.~\ref{Fig3} and \ref{Fig4}{\bf a}, we use the same setup as above and record a video with microspheres (silica, diameter $1.98\,{\rm \mu m}$, Microparticles GmbH) diffusing above a coverslip under the same illumination conditions as above. 
These experiments are performed at $18\,{\rm fps}$ with a field of view of $500 \times 500\,{\rm px}$ ($35\times35\,{\rm \mu m}$) for Fig.~\ref{Fig3} and $400 \times 400\,{\rm px}$ ($28\times28\,{\rm \mu m}$) for Fig.~\ref{Fig4}{\bf a}.

For the multiple-particle experiments shown in Figs.~\ref{Fig4}{\bf b}-{\bf c}, we use a fluorescence microscopy setup to record fluorescent bacteria (\emph{Bacillus subtilis} wild type strain NCIB 3610, see below) swimming above a coverslip without (Fig.~\ref{Fig4}{\bf b}) and with (Fig.~\ref{Fig4}{\bf c}) microsphere (silica, diameter $4.28\,{\rm \mu m}$, Microparticles GmbH) in the solution.
The sample is illuminated with LEDs (Thorlabs LED MCWHL5, white light, and M490L4, blue light for fluorescent illumination) and the video is recorded with a sCMOS camera (Hamamatsu, C14440-20UP) using a field of view of $250 \times 250\,{\rm px}$ (corresponding to $65 \times 65\,{\rm \mu m}$) at $20\,{\rm fps}$.
Due to the low intensity of the fluorescence emitted by the bacteria, the gain of the sCMOS camera is increased leading to a high level of electronic noise. 

\textbf{Equipartition method analysis.} The probability distribution of a particle position can be obtained from the histogram of the particle position in a sufficiently long trajectory. In an optical trap, this probability distribution is typically a Gaussian function \cite{jones2015optical}:
\begin{equation}
p(x) \propto \exp\left(-\frac{x^2}{2\sigma^2_x}\right),
\end{equation}
where $x$ is the particle position with respect to the trap center and $\sigma^2_x$ is the particle position variance. 
For the optically trapped particle in Fig.~\ref{Fig2}, $\sigma^2_x= 0.89\,{\rm px^2}$.

\textbf{Auto-correlation function analysis.} 
The auto-correlation function (ACF) can be calculated from a particle trajectory. For an optically trapped particle, the ACF is \cite{jones2015optical}
\begin{equation}
\mathcal{C}_x(\tau) = \exp\left(-\frac{\tau}{\tau_x}\right)
\end{equation}
where $\tau_x$ is the trap characteristic time.
For the optically trapped particle in Fig.~\ref{Fig2}, $\tau_x = 20\,{\rm ms}$.

\textbf{Bacteria preparation.} 
The \emph{Bacillus subtilis} wild type strain NCIB 3610 is optimized for motility before conducting the experiments \cite{wolfe1989migration, pincce2016disorder}. The \emph{B. subtilis} cell culture is prepared by seeding cells from a frozen stock in lysogen broth (LB) medium (Miller, Sigma-Aldrich) and growing them overnight at $37^{\circ}{\rm C}$, shaking at $180\,{\rm rpm}$ (1 colony in $5\,{\rm ml}$ of LB growth medium in a $50\,{\rm ml}$ falcon tube). The saturated cell culture is then diluted (1:100) in fresh LB medium and grown for another $4.5\,{\rm h}$, until the culture reaches the mid-exponential phase of bacterial growth corresponding to the state where the cells are most motile. A $5\,{\rm \mu l}$ aliquot of the diluted culture is inoculated in a semi-solid agar plate consisting of LB medium with $0.5\%{\rm w/v}$ agar (Merck) and incubated at $30^{\circ}{\rm C}$ overnight. The motile cells move outwards from the point of inoculation allowing for the selection of the most motile cells by collecting cells from the leading edge of the ring. These cells are then cultured in LB medium overnight under the same conditions as before and a sample of this culture, with $15\%{\rm (v/v)}$ glycerol, is frozen at $-80^{\circ}{\rm C}$ to keep as a stock of motile \emph{B. subtilis}. For each experimental session, a cell culture is grown from the motile stock, overnight and for $4.5\,{\rm h}$ the next day, as before. Next, the cells are stained using the Live/Dead BacLight Bacterial Viablility Kit (Thermo Fisher) for fluorescent imaging. Since the staining is only to be able to readily distinguish between bacterial cells and microparticles in the analysis, the cells were only stained with the SYTO 9 stain by adding $1\,{\rm \mu l}$ to $2\,{\rm ml}$ of the cell culture and incubate further for $30\,{\rm min}$. The SYTO 9 stain is a green-fluorescent nucleic acid stain with excitation/emission maxima at $480$ and $500\,{\rm nm}$ and, when used alone, stains all bacterial cells. Finally, the fluorescent cell culture is washed two times by centrifuging the culture at $4500\,{\rm g}$ for $4\,{\rm min}$ and resuspended in motility buffer ($0.01\,{\rm M}$ potassium phosphate (KH2PO4), $0.1\,{\rm mM}$ EDTA, $0.01\,{\rm M}$ glucose, $0.002\%$ Tween-20, ${\rm pH}\,7.0$) \cite{adler1967effect}. 

\textbf{Codes.}
We provide a Python freeware software package, called DeepTrack, which can be readily personalized and optimized for the needs of specific users and applications.
The key functions are contained in the library ``deeptrack.py", 
while five applications are provided in the Example Codes 1-4, each of which consists of a Jupyter notebook and some additional files for the pre-trained neural networks (in .h5 format) and some sample videos (in .mp4 format).

Example Code 1a generates some particle images like those employed in Figs.~\ref{Fig1}{\bf b}-{\bf c} and tracks their position using the pre-trained network ``DeepTrack - Example 1a - Pretrained network.h5".

Example Code 1b demonstrates the whole training process for the neural network employed in Figs.~\ref{Fig1}{\bf b}-{\bf c}: from definition of the parameters for the simulation of the particle images, to the definition of the neural network, to its training, and finally to its testing. 

Example Code 2 generates some particle images like those employed to train the neural network used in Fig.~\ref{Fig2}.
Furthermore, using the pre-trained neural network ``DeepTrack - Example 2 - Pretrained network.h5", it tracks the positions of the optically trapped particle in ideal (see Figs.~\ref{Fig2}{\bf a}-{\bf d} and Video 1) and poor (see Figs.~\ref{Fig2}{\bf e}-{\bf h} and Video 2) illumination conditions using the corresponding original videos (``DeepTrack - Example 2 - Optically Trapped Particle Good.mp4" and ``DeepTrack - Example 2 - Optically Trapped Particle Bad.mp4", respectively).

Example Code 3 generates some particle images like those employed to train the neural network used in Figs.~\ref{Fig3} and \ref{Fig4}{\bf a}-{\bf b}.
Furthermore, using the pre-trained neural network ``DeepTrack - Example 3 - Pretrained network.h5", it tracks the positions of multiple particles in ideal (see Figs.~\ref{Fig3} and Video 3) and poor (see Fig.~\ref{Fig4}{\bf a} and Video 4) illumination conditions as well as bacteria (see Fig.~\ref{Fig4}{\bf b} and Video 5) using the corresponding original videos (``DeepTrack - Example 3 - Brownian Particles Good.mp4", ``DeepTrack - Example 3 - Brownian Particles Bad.mp4", and ``DeepTrack - Example 3 - Bacteria.mp4", respectively).

Example Code 4 generates some particle images like those employed to train the neural network used in Fig.~\ref{Fig4}{\bf c}.
Furthermore, using the pre-trained neural network ``DeepTrack - Example 4 - Pretrained network.h5", it tracks the positions of multiple particles in poor illumination conditions and in the presence of bacteria (see Fig.~\ref{Fig4}{\bf c} and Video 6) using the corresponding original video (``DeepTrack - Example 4 - Brownian particles + bacteria.mp4").



We thank Falko Schmidt for help with the figures, Giorgio Volpe for critical reading of the manuscript, and Santosh Pandit and Er\c{c}a\u{g} Pin\c{c}e for help and guidance with the bacterial cultures.
%
This work was supported by the ERC Starting Grant ComplexSwimmers (Grant No. 677511).
%
%
%

\end{document}